\documentclass[a4paper,11pt]{article}

\usepackage{jinstpub} 

\usepackage[utf8]{inputenc}
\usepackage{siunitx}
\usepackage{multirow}
\usepackage{graphicx}
\usepackage{mathrsfs}
\graphicspath{ {./figures/} }
\usepackage{caption}
\usepackage{subcaption}
\usepackage{color}

\usepackage{lineno}

\title{\boldmath The level-1 trigger for the SuperCDMS experiment at SNOLAB}

\author[a,1]{J.~S.~Wilson,\note{Corresponding author.}}
\author[b,c,1]{H.~Meyer~zu~Theenhausen,}
\author[b,c]{B.~von~Krosigk,}
\author[d]{E.~Azadbakht,}
\author[e]{R.~Bunker,}
\author[f,g]{J.~Hall,}
\author[h]{S.~Hansen,}
\author[i]{B.~Hines,}
\author[e]{B.~Loer,}
\author[h]{J.~T.~Olsen,}
\author[j,k]{S.~M.~Oser,}
\author[l]{R.~Partridge,}
\author[m,n]{M.~Pyle,}
\author[o]{J.~Sander,}
\author[m]{B.~Serfass,}
\author[d]{D.~Toback,}
\author[m]{S.~L.~Watkins,}
\author[d]{X.~Zhao}

\affiliation[a]{Baylor University,\\Waco 76706, TX, USA}
\affiliation[b]{Institute for Astroparticle Physics (IAP), Karlsruhe Institute of Technology (KIT),\\76344 Eggenstein-Leopoldshafen, Germany}
\affiliation[c]{Institut f{\"u}r Experimentalphysik, Universit{\"a}t Hamburg\\22761 Hamburg, Germany}
\affiliation[d]{Department of Physics and Astronomy, and the Mitchell Institute for Fundamental Physics and Astronomy, Texas A\&M University, \\College Station, TX 77843, USA}
\affiliation[e]{Pacific Northwest National Laboratory, \\Richland, WA 99352, USA}
\affiliation[f]{SNOLAB,\\
Sudbury, ON P3Y 1N2, Canada}
\affiliation[g]{Laurentian University, Department of Physics,\\ 
Sudbury, Ontario P3E 2C6, Canada}
\affiliation[h]{Fermi National Accelerator Laboratory, \\Batavia, IL 60510, USA}
\affiliation[i]{Department of Physics, University of Colorado Denver, \\Denver, CO 80217, USA}
\affiliation[j]{Department of Physics \& Astronomy, University of British Columbia, \\Vancouver, BC V6T 1Z1, Canada}
\affiliation[k]{TRIUMF, \\Vancouver, BC V6T 2A3, Canada}
\affiliation[l]{SLAC National Accelerator Laboratory/Kavli Institute for Particle Astrophysics and Cosmology, \\Menlo Park, CA 94025, USA}
\affiliation[m]{Department of Physics, University of California, \\Berkeley, CA 94720, USA}
\affiliation[n]{Lawrence Berkeley National Laboratory, \\Berkeley, CA 94720, USA}
\affiliation[o]{Department of Physics, University of South Dakota, \\Vermillion, SD 57069, USA}

\emailAdd{Jon\_Wilson2@baylor.edu, hanno.theenhausen@kit.edu}

\abstract{
    The SuperCDMS SNOLAB dark matter search experiment aims to be sensitive to energy depositions down to $\mathcal{O}$(1\,eV). This imposes requirements on the resolution, signal efficiency, and noise rejection of the trigger system. To accomplish this, the SuperCDMS level-1 trigger system is implemented in an FPGA on a custom PCB. A time-domain optimal filter algorithm realized as a finite impulse response filter provides a baseline resolution of 0.38 times the standard deviation of the noise, $\sigma_{\rm n}$, and a 99.9\% trigger efficiency for signal amplitudes of 1.1 $\sigma_{\rm n}$ in typical noise conditions. Embedded in a modular architecture, flexible trigger logic enables reliable triggering and vetoing in a dead-time-free manner for a variety of purposes and run conditions. The trigger architecture and performance are detailed in this article.
}


\begin{document}
\maketitle
\flushbottom

\section{Introduction}
\label{sec:intro}
\begin{figure}[b]
    \centering
    \includegraphics[width=0.95\textwidth]{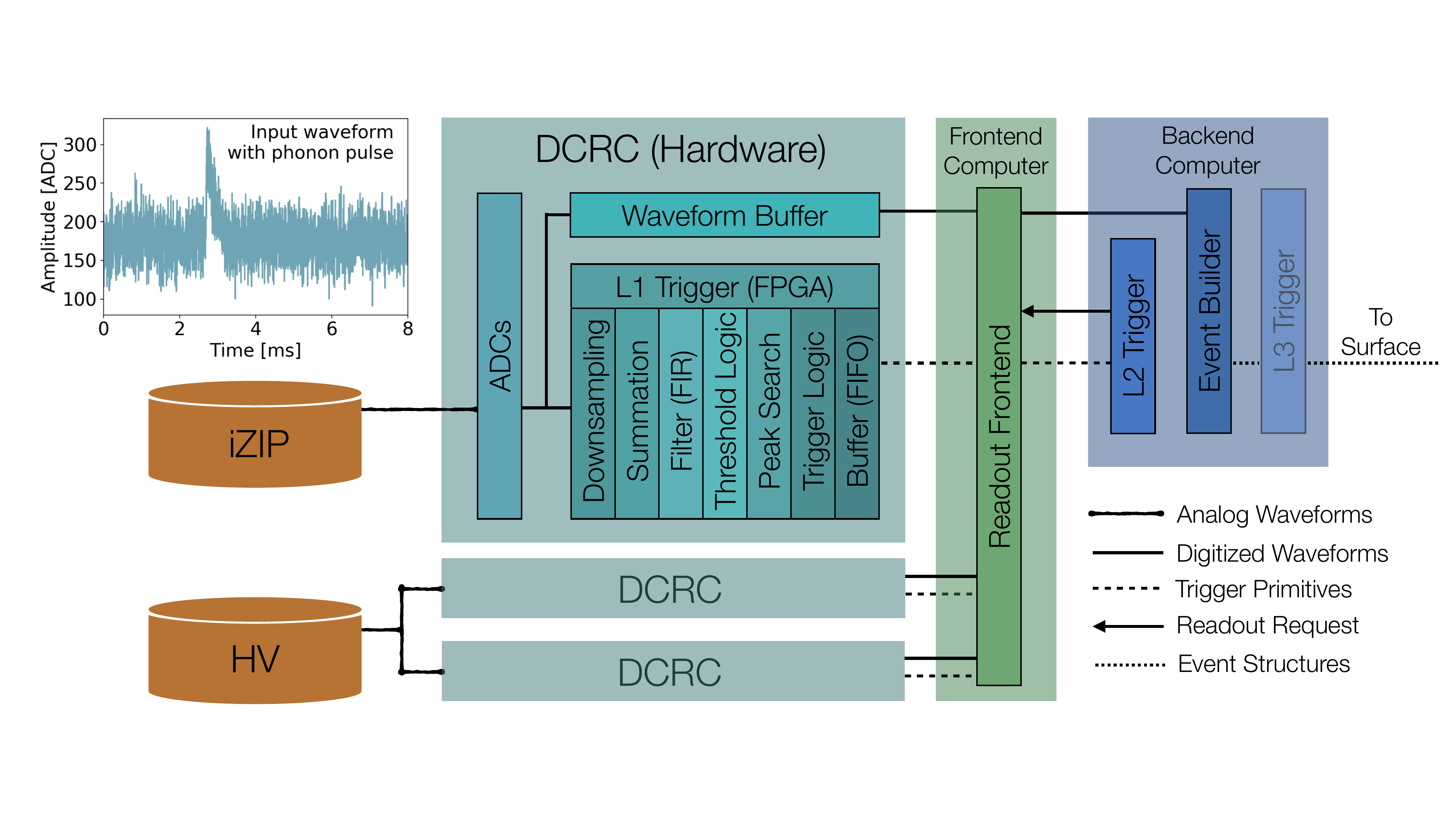}
    \caption{Schematic overview of the SuperCDMS trigger and data acquisition system. Input waveforms from the iZIP and HV detectors are each read out by one or two custom Detector Control and Readout Cards (DCRCs), respectively, which include the level-1 trigger. In parallel to the decision-making within the L1 trigger, the digitized waveforms are stored for up to \SI{6.71}{\second} in a circular ``waveform buffer''. The DCRCs are controlled by custom software on frontend computers. The level-2 and the optional level-3 trigger combine the information from all detectors and are implemented in software running on a backend computer.}
    \label{fig:daqtriggerschematic}
\end{figure}
The SuperCDMS experiment at SNOLAB is a direct dark matter search experiment based on cryogenic crystal detectors made from germanium and silicon~\cite{Agnese_2017}.
Particle interactions within the detector produce phonon and ionization signals that can be measured using superconducting transition edge sensors and charge electrodes, respectively.
One detector type, the ``iZIP'' detector\footnote{iZIP detector: interleaved Z-sensitive Ionization Phonon detector.}, employs both sensor types, which allows the experiment to discriminate between nuclear and electron recoils.
Another detector type, the high voltage (HV) detector, omits the charge sensors, and instead applies a high bias voltage resulting in a large Neganov-Trofimov-Luke amplification of the phonon signal~\cite{Neganov:1985khw,Luke1988VoltageassistedCI}.
The HV detector has increased sensitivity towards lower energy signals, but the discrimination between nuclear and electron recoils is lost.
Using both detector types, SuperCDMS is designed to detect electron-equivalent recoil energies ranging from $\mathcal{O}(\SI{100}{\kilo\eV})$ down to $\mathcal{O}(\SI{1}{\eV})$.
At the lowest energies, the detector sensitivity is limited by vibrational and electronic noise.
Improvement in the noise levels and in the ability to extract signals from the noise is critical to the science goals of SuperCDMS.

Recording all of the data from the detectors, which operate continuously, would overburden the bandwidth and storage capacity of the data acquisition system.
For this reason, a trigger system is used to record only a small portion of the data for offline analysis.
Realizing the SuperCDMS science goals requires high-efficiency acquisition at the very lowest energies where event rates due to backgrounds and noise tend to be higher.
To achieve this, SuperCDMS uses a three-stage trigger system schematically shown in figure~\ref{fig:daqtriggerschematic}, consisting of both hardware- and software-level stages, that makes real time decisions about which portions of the data are to be recorded. 

In this manuscript we describe the level-1 (L1) trigger. The L1 trigger only uses data from a single detector, while the level-2 (L2) trigger and the optional level-3 trigger combine L1 trigger information from all detectors.
The trigger system operates during both low-rate dark matter search periods and high interaction rate calibration periods, and operates on both iZIP and HV detectors.
It provides robust and efficient triggering above an energy threshold that is as low as achievable while remaining dead-time-free.
Digitized input traces are analyzed in a complex L1 trigger architecture, described in section~\ref{sec:arch}, with a finite-impulse-response (FIR) filter at its heart. The design of an optimal filter (OF) FIR is introduced in section~\ref{sec:theo}, which also describes the pathological ``echo triggers" that arise with the use of this filter. 
Section~\ref{sec:QualitativePerformance} details the qualitative performance of the OF FIR and how echo triggers can be circumvented by combining the OF with a boxcar filter (BF) in the trigger logic when it is needed. The OF design is compared quantitatively to a matched filter (MF) and BF in section~\ref{sec:QuantitativePerformance} before the concluding section~\ref{sec:conclusion}.

\section{L1 Trigger Architecture}
\label{sec:arch}
\begin{figure}[b!]
    \centering
    \includegraphics[width=1\textwidth]{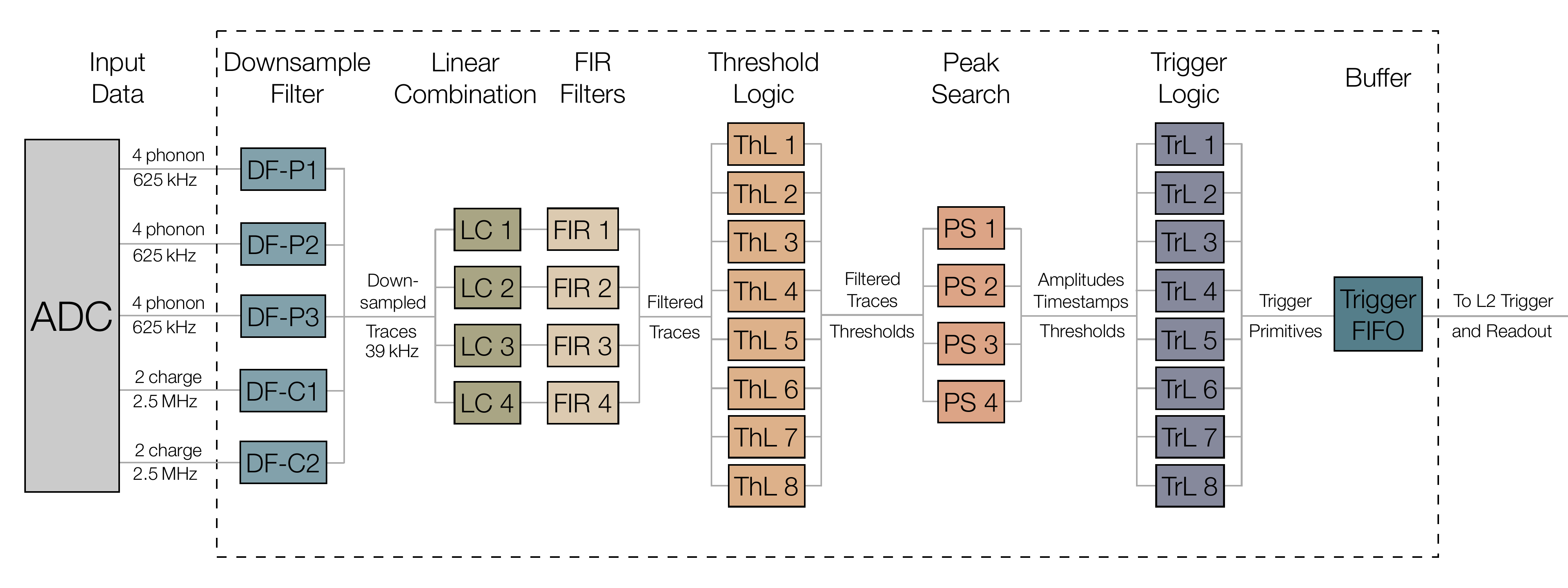}
    \caption{
    Schematic overview of the SuperCDMS L1 trigger architecture consisting of several discrete modules. The dashed line indicates the L1 trigger on the FPGA which receives inputs from the ADC on the same DCRC. The Linear Combination, FIR and Peak Search modules, with a configurable Threshold Logic in between, are replicated four times to produce parallel trigger paths whose outputs can be combined with one another in the final Trigger Logic.
    The elements and acronyms shown in this figure are described in sections~\ref{sec:input} through~\ref{sec:FIFO}.
    }
    \label{fig:triggerarch}
\end{figure}

The trigger system consists of several discrete modules which are described in this section and depicted in figure~\ref{fig:triggerarch}. It is designed to make its trigger decision based on an estimate of the energy of an interaction in the SuperCDMS detectors. The energy is expected to be linearly related to the amplitude of the resulting detector pulse,
thus any linear filter applied to the data can be interpreted as providing a direct estimate of the energy of a hypothetical pulse occurring at each point in time. 

The L1 trigger is designed around a flexible FIR filter, described in section~\ref{sec:FIR}, which can be optimized to provide the most precise energy estimate possible.
The FIR is preceded by data preconditioning modules (sections~\ref{sec:downsample} and \ref{sec:linearcomb}), and followed by modules that identify and characterize pulses in the FIR output (sections~\ref{sec:ThL} and \ref{sec:peaksearch}).
The preconditioning, FIR, and pulse identification and characterization modules are replicated four times to produce independent and parallel trigger paths.
Up to eight simple Boolean combinations of the results of these four trigger paths are then computed (section~\ref{sec:TrL}) to produce the final L1 trigger results, which are stored in a first-in, first-out (FIFO) buffer (section~\ref{sec:FIFO}) for retrieval and use by the L2 trigger.
This L1 trigger architecture, as designed, provides a high degree of flexibility to adapt to experimental conditions and needs while maintaining very low experimental thresholds and providing zero dead time.

The modules are implemented in firmware running on an Altera Cyclone IV FPGA~\cite{cycloneIV} on the Detector Control and Readout Card (DCRC).
Data is transferred from one module to the next using the Altera Avalon-ST protocol, while run time configuration and readout are handled via the Avalon-MM protocol~\cite{Avalon-spec}.
The design and implementation have been tested and shown to perform as expected in a variety of ways, as described in section~\ref{sec:validation}.

\subsection{Input Data}
\label{sec:input}

The incoming data to each DCRC's L1 trigger consists of twelve phonon channels and four charge channels \cite{Agnese_2017}, which constitute all the channels from one iZIP detector.
The L1 trigger for an HV detector only receives data from channels belonging to a single side of the detector; the other trigger input channels are ignored.
The ADCs convert the analog inputs to 16-bit unsigned integers; the sample rate is \SI{625}{\kilo\Hz}.
The samples from the charge channels are also 16-bit unsigned integers, but the sample rate is \SI{2.5}{\mega\Hz}.
    
\subsection{Downsample Filter}
\label{sec:downsample}

The first step is to bring the sample rate of all channels down to a common optimal level and to synchronize them. 
We employ a cascaded integrator-comb (CIC) low-pass filter, which is a generalized moving-average filter, to mitigate aliasing of high-frequency noise when downsampling~\cite{CIC_filter}.
The charge channels are downsampled by a factor of 64, and the phonon channels are downsampled by a factor of 16, yielding a final sample rate for all channels of \SI{39.0625}{\kilo\Hz}.
The transfer function of a CIC filter, which relates the output amplitude at every frequency to the input amplitude at the same frequency, is
\begin{equation}
    H_\text{CIC}(z) = \left(\frac{1 - z^{-RM}}{1 - z^{-1}}\right)^N,
\end{equation}
where $z$ is the complex variable of the $z$-transform~\cite{z-transform}, $R$ is the downsampling factor of 64 (16) for charge (phonon) channels, $M$ is the number of samples per stage, and $N$ is the order of the filter.
We choose $R = 64$ (16), $M = 1$, and $N = 3$ based on studies optimizing the sensitivity of the trigger, similar to those described in section~\ref{sec:theo}.

To synchronize the individual channels, incoming samples are stored in holding-area registers and sent out only when the corresponding samples of all sixteen channels have arrived.
After this point, the L1 trigger no longer makes a distinction between phonon and charge channels because the timing and sample rates of all sixteen channels are identical.

\subsection{Linear Combination}
\label{sec:linearcomb}

Each of the four trigger paths begins with a configurable linear combination of the input channels.
This gives the L1 trigger the flexibility to implement common triggering schemes, including the sum of all the phonon channels, the sum of the inner charge channels, or the sum of the phonon channels on one side, as well as less common possibilities such as the sum of phonon and charge channels. Additionally, the linear combination allows calibrations to be applied to each channel individually to accommodate differences in sensitivity from one channel to another.

This is implemented by multiplying each channel's data by an 8-bit signed coefficient, configurable at run time.
Then the sixteen channels are summed together. Each of the four trigger paths contains a linear combination module that is independent of the other paths.

\subsection{Finite Impulse Response}
\label{sec:FIR}

The heart of the L1 system and each trigger path is the FIR filter.
The FIR filter can be used to implement the OF described in section~\ref{sec:theo}, as well as a number of other possibilities, which are discussed in section~\ref{sec:perf}.
An infinite impulse response filter could also be used to implement these filter schemes and more, but we have chosen to use an FIR filter because it is simpler to implement and is inherently stable~\cite{fir_stability}.

Each of the four FIR filters provides an instantaneous estimate of the energy.
This is accomplished by computing a sum over the 1024 most recent input samples, each weighted by its own coefficient.
This sum spans a time window of \SI{1.6}{\milli\s}.
The sequence of coefficients, each of which is a 16-bit signed integer, is stored in the FIR.
As each sample is received from the Linear Combination module, it is stored in a FIFO buffer that is 1024 samples long.
After each sample arrives, the FIR module iterates over the FIFO and the stored coefficients, multiplying each sample by the corresponding coefficient and summing the result.
Thus the $n$th sample $x[n]$ received by the FIR produces one filtered output sample $y[n]$, given by
\begin{equation}
\label{eq:FIRarch}
    y[n]=\sum_{i=0}^{1023} b_i x[n-i] , \qquad H_\text{FIR}(z) = \sum_{i=0}^{1023} b_i z^{-i},
\end{equation}
where $b_i$ is the $i$th coefficient and $H_\text{FIR}(z)$ is the transfer function of this filter, with $z$ being the complex variable of the $z$-transform.

Because the FIR is a linear filter, as are the preconditioning steps, each $y[n]$ represents an instantaneous estimate of the amplitude, and therefore the energy, of a putative pulse occurring at the time of sample $n$.
The filter design determines the degree to which noise in the input is suppressed, and therefore the resolution of the energy estimate and the performance of the trigger.

The downsample filter, linear combination, and FIR modules each increase the bit width of the data, and at this point, the bit width of the data has grown large enough (72 bits) that we need to reduce it.
The FIR output samples are shifted left by a configurable number of places, which corresponds to multiplication by a power of two, and then the 40 least-significant bits are truncated, leaving 32 bits.
If the left-shifted value exceeds the maximum value (or most negative value) that is representable using 32 bits, then it is replaced by the maximum value (or most negative value) possible.

\subsection{Threshold Logic}
\label{sec:ThL}

The Threshold Logic modules compare the associated FIR outputs to configurable thresholds.
They each record their results as a single-bit, which identifies potentially-significant energy deposits to be further examined in the Peak Search modules.

When the FIR output first rises above an activation threshold the threshold output changes to 1.
To mitigate against re-triggering caused by noise, the output remains at 1 until the FIR output falls back below a lower deactivation threshold, which is also configurable, at which point the output changes back to 0.
The output of all eight threshold modules is conveyed along with the four FIR outputs to the peak search modules.

While the simplest threshold logic scheme has a single threshold module for each trigger path, we note that each of eight threshold modules may be applied to any one of the four FIR outputs.
This flexibility allows complex triggering schemes to be used if necessary, for example enabling rejection of high-energy pulses that might otherwise overwhelm the data acquisition system while retaining low-energy pulses during source calibrations.

\subsection{Peak Search}
\label{sec:peaksearch}

The peak search module uses the FIR and Threshold Logic results to produce peak time, amplitude and threshold information to be used by the Trigger Logic module to make the final L1 decision.
Each of the four trigger paths has a corresponding peak search module that receives the output of the FIR in the same trigger path.
The peak search module also receives the output of all eight threshold modules regardless of which of the four trigger paths they are associated with.

The ``trigger window'' is the continuous period of time during which any of the threshold modules associated with one FIR are activated.
Within the trigger window, the peak search module finds and records the maximum FIR output and the timestamp at which that maximum occurred.
Additionally, the peak search module records the state of all eight thresholds at the time of the peak, and it records which of the eight threshold bits were 1 at any point during the entire trigger window.

At the end of the trigger window, the peak search module creates the ``trigger primitives''.
These consist of the maximum FIR value (32 bits), the timestamp at which the maximum occurred (32 bits), the state of the thresholds at the time of the peak (8 bits), and which thresholds were active at any time during the entire trigger window (8 bits).
At other times, the peak search block does not produce any outputs.

In special cases where very high-energy events produce pulses that saturate the ADC value, and last longer than lower-energy pulses (``saturated pulses''), the peak search module provides an alternative handling to determine the peak time. Specifically, if the duration of the trigger window has exceeded a configurable threshold, then instead of recording the time at which the maximum occurred, the timestamp included in the trigger primitives will be at a configurable time after the trigger window began.

\subsection{Trigger Logic}
\label{sec:TrL}

The Trigger Logic modules produce the final L1 trigger decisions based on configurable logical combinations of multiple threshold states from the trigger primitives of an associated trigger path.
The accepted trigger primitives are then stored in a FIFO buffer for readout.

Every time one of the four peak search modules produces a set of trigger primitives, the trigger logic module examines the 16 bits that describe the state of the thresholds at both the time of the peak and during the entire trigger window.
Depending on which of the four trigger paths produced the trigger, 
each of the 16 bits may be required to be 1, required to be 0, or ignored.
These logical requirements are configurable at run time.
In addition to the requirements placed on the threshold-state bits, triggers may be rejected with a configurable probability, or prescale.

The L1 trigger system includes 8 Trigger Logic modules, each providing one set of logical requirements and prescales.
A trigger is accepted as long as it satisfies at least one of these 8 sets of logical requirements and prescales.
The results of the 8 trigger decisions are appended to the trigger primitives as an additional 8 bits, where 1 indicates acceptance and 0 indicates rejection. 
All passing primitives are stored in the trigger FIFO for later readout by the data acquisition system.

This architecture gives the L1 trigger the flexibility to define trigger decisions in terms of multiple trigger paths.
For example, two trigger paths could be configured to use disjoint sets of channels, and then the trigger decision could require both trigger paths to exceed a threshold simultaneously.
This scheme would reject spurious triggers caused by uncorrelated noise. Section~\ref{sec:QualitativePerformance} provides a case study in which the information from two trigger paths is combined in the trigger logic to reject spurious ``echo triggers'' related to the OF algorithm.

\subsection{Trigger FIFO and Vetoes}
\label{sec:FIFO}

The Trigger FIFO and Vetoes system is designed to store trigger primitives that are accepted by the Threshold Logic module for readout by the data acquisition system and also to measure and record the live time of the L1 trigger. 
The trigger primitives are stored in a FIFO data structure, from which the data acquisition system can read the triggers.
This FIFO buffer can hold up to 256 triggers.

In the event that the buffer is full and cannot accept more triggers, an entry is made in a secondary FIFO buffer, the ``veto FIFO'', indicating the time at which the full condition began.
When the buffer is no longer full because some triggers have been read out by the data acquisition system, another entry is made in the veto FIFO indicating the time at which the full condition ended.
Although the trigger system is designed to gracefully handle and record this dead time if it ever occurs, the data acquisition system is designed to read the buffer frequently enough to maintain dead-time-free operation.

Additionally, veto periods can be requested by external sources, for example in response to the environmental conditions of the experiment. The beginning and end of these externally-requested veto periods are recorded into the veto FIFO.
Like the trigger FIFO, the veto FIFO can be read and emptied by the data acquisition system.
If the veto FIFO becomes over-full, then an error is recorded to indicate that information has been lost.

Triggers that are lost during veto periods are counted, and the count is stored for readout.
The total amount of time during which triggers are able to be accepted, the live time, and the total amount of time during veto periods, the veto time, are tracked and also available for readout.

The total L1 trigger latency, or the elapsed time between an event and the entry of the corresponding trigger in the Trigger FIFO, is $\sim\SI{26}{\milli\second}$.
This latency is dominated by the duration of the trigger window.
The data acquisition system involves much larger latencies, $\mathcal{O}(\SI{1}{\second})$, so the L1 trigger latency is negligible.

\subsection{Functional Validation}
\label{sec:validation}

The functioning of the trigger system shown in figure~\ref{fig:triggerarch} has been validated in several different ways to single bit precision. 
Each component of the trigger is tested in isolation using logic simulation software~\cite{Quartus-ModelSim} (simulation test bench) and in the same FPGA~\cite{cycloneIV} as is used for the complete trigger system (hardware test bench). To test the logic, a sequence of inputs was devised to exercise every path in the design in both the simulation and hardware test benches.
The expected outputs were calculated by hand and compared to the test bench outputs.

Finally, a full trigger simulation software suite was developed and validated.
It reproduced the results from the same sequence of inputs.
In addition, it was fed the raw data from real data taking at the SuperCDMS detector testing facility at SLAC from which it reproduced the outputs of the FPGA implementation of the trigger. This software package will be used in online monitoring of the experiment and applied to simulated events to estimate the efficiency of the triggers and analyze the experimental data.

\section{Filter Configuration}
\label{sec:theo}
Because the goal is to optimize the signal to noise rejection at as low an energy threshold as possible, we have chosen to implement a time-domain optimal filter method. In this section we outline the OF formalism and implementation within the FIR filter described in section~\ref{sec:FIR}, and explain the methods for choosing the FIR coefficients for various scenarios. We show that the OF response to large pulses can lead to pathological echo triggers, and we compare the expected OF resolution and response to those expected for both a matched filter (MF) \cite{matched} and a boxcar filter (BF) \cite{boxcar}.
A filter configuration with high resolution and without echo triggers, created via a combination of the OF and the BF, is described in section~\ref{sec:QualitativePerformance}.

\subsection{Optimal Filter FIR}
The OF is used as a minimum variance estimator of the amplitude of a pulse, with a known shape, in the presence of stationary noise \cite{OF_gatti}. The OF, $\phi(t)$, in the time domain, $t$, is derived from a $\chi^2$ equation in the frequency domain, $\nu$. The $\chi^2$ equation relates a general signal $S(\nu)$ to a pulse template $A(\nu)$ scaled by the amplitude $a$, and weighted by the noise power spectral density (PSD) $J(\nu)$, given by\footnote{In the discrete notation of eq.~\ref{eq:FIRarch}, the final results $\phi(t)$, $S(t)$ and $a$ correspond to $b_i$, $x[n-i]$ and $y[n]$.}:
\begin{equation}
    \chi^2(\nu; a) = \sum_{\nu} \frac{\left| S(\nu) - a A(\nu) \right|^2}{J(\nu)}.
\end{equation}
From the minimum variance we find
\begin{equation}
    a = \sum_{\nu} \phi(\nu) S(\nu) , \qquad \phi(\nu) = \frac{A^*(\nu)/J(\nu)}{\sum_{\nu'}A(\nu')A^*(\nu')/J(\nu')},
\end{equation}
where $\phi(\nu)$ is the frequency-domain OF.
The estimator for the time-domain OF, $\phi(t)$, is given by inverse Fourier transformation with
\begin{equation}
    a = \sum_{t} \phi(t) S(t) , \qquad \phi(t) = \mathscr{F}^{-1}\left( \frac{A^*(\nu)/J(\nu)}{\sum_{\nu'}A(\nu')A^*(\nu')/J(\nu')}  \right).
\label{eq:FIReq}
\end{equation}

In order to calculate the OF FIR coefficients for the L1 trigger, channel-specific PSDs measured in data and expected templates first need to be propagated through the respective trigger paths up to the FIR module using the trigger simulation. The resulting PSDs and templates are then used according to eq.~(\ref{eq:FIReq}) to calculate the OF FIR coefficients, represented by each entry of $\phi(t)$. Additionally, the coefficients are scaled to span the maximum coefficient bit range and overall shifted to ensure a zero DC component (the coefficients sum to 0). 

The expected FIR filter baseline resolution, $\sigma_{\text{FIR}}$, and response to a signal template, $R$, (proportional to the amplification) are calculated as
\begin{equation}
    \sigma_\text{FIR} = \sqrt{\sum_{\nu} J(\nu) \left|\phi(\nu) \right|^2} , \qquad R = \sum_{t} \phi(t) A(t).
\label{eq:resolution}
\end{equation}
The ratio of response and resolution provides an estimator for the signal-to-noise ratio.

\subsection{Filter Configuration in Different Scenarios}

\begin{figure}[b!]
     \centering
     \begin{subfigure}[b]{0.495\textwidth}
         \centering
         \includegraphics[width=\textwidth]{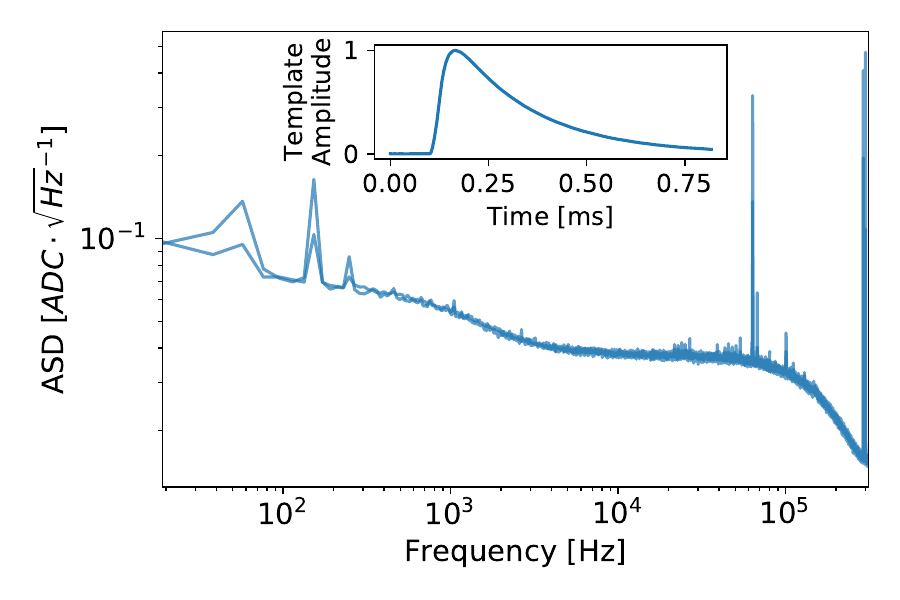}
         \caption{}
         \label{fig:PSD_TB}
     \end{subfigure}
     \hfill
     \begin{subfigure}[b]{0.495\textwidth}
         \centering
         \includegraphics[width=\textwidth]{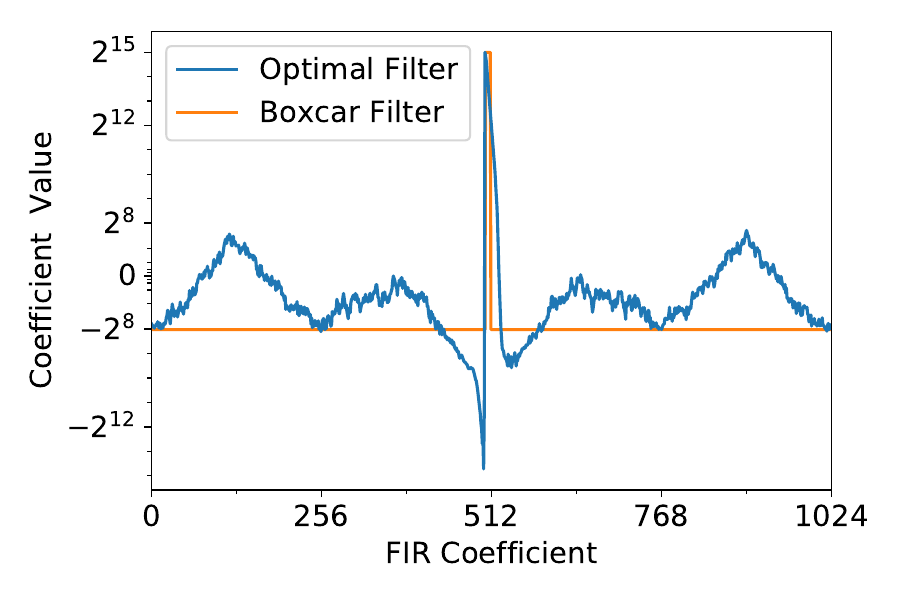}
         \caption{}
         \label{fig:FIR_COEFFS_TB}
     \end{subfigure}
        \caption{(a) Amplitude spectral density (ASD) of two DCRC channels on a test bench setup with no detector attached. The inset shows the pulse template which is used in the arbitrary waveform generator and to derive the FIR coefficients. (b) FIR coefficients of an optimal filter and a boxcar filter.}
        \label{fig:TB_PSD_FIR}
\end{figure}

In this subsection we show the computed coefficients for the OF, MF and BF for different noise scenarios, and explain how echo triggers can arise when using the OF.
A baseline set of conditions are shown in figure~\ref{fig:TB_PSD_FIR} (a). It shows the square roots of the noise PSDs (the amplitude spectral densities or ASDs) from two DCRC channels in a test bench setup with no detector attached, as well as templates from averaged arbitrary waveform generator (AWG) input pulses to these channels. The ASDs trend to higher noise contributions at lower frequencies with a few large spikes at high and low frequencies attributed to the AWG and the power supply frequency. 
The phonon templates have a typical double exponential shape. The resulting OF FIR coefficients from these PSDs and templates, alongside the coefficients from a BF, are displayed in figure~\ref{fig:TB_PSD_FIR} (b). 
The OF coefficients feature a sharp central peak surrounded by noise-mitigating baseline oscillations.
The low frequency spikes in the noise PSDs cause oscillations up to positive values in the corresponding OF FIR coefficients. 

Very large pulses can be amplified by these oscillations to have an amplitude that is above the trigger threshold and cause artifact ``echo triggers'' close to the main trigger.
The number of expected echo triggers depends on the FIR window length in relation to the period of the responsible noise components.
In this case two echo triggers are expected, one preceding and one succeeding the primary trigger. Since the MF and BF do not have positive FIR coefficients outside the peak region, they do not produce echo triggers. In section \ref{sec:QualitativePerformance} it is demonstrated how echo triggers can be circumvented in the trigger logic by combination with the BF. With a box width of 5 samples the BF coefficients resemble the best fit to the template.

\begin{figure}[t!]
     \centering
     \begin{subfigure}[b]{0.495\textwidth}
         \centering
         \includegraphics[width=\textwidth]{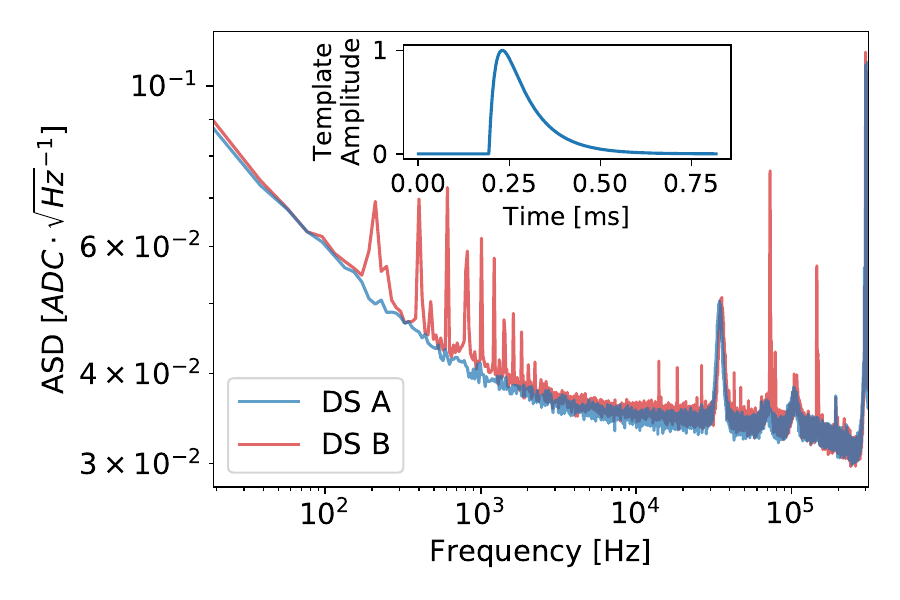}
         \caption{}
         \label{fig:PSD_G124}
     \end{subfigure}
     \hfill
     \begin{subfigure}[b]{0.495\textwidth}
         \centering
         \includegraphics[width=\textwidth]{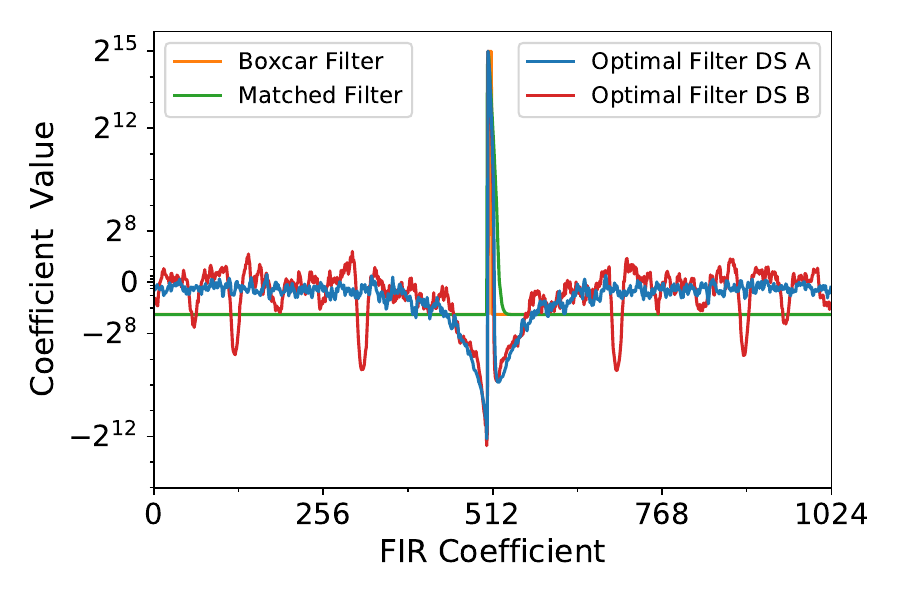}
         \caption{}
         \label{fig:FIR_COEFFS_G124}
     \end{subfigure}
        \caption{(a) Amplitude spectral density (ASD) and pulse template of a single channel from two datasets (DS A and DS B) with different noise conditions, taken with a SuperCDMS SNOLAB prototype detector. (b) FIR coefficients of two optimal filters optimized to the respective datasets, a matched filter and a boxcar filter. Since the pulse templates in both datasets are identical, the matched filter and boxcar filter are optimized to both.}
        \label{fig:G124_PSD_FIR}
\end{figure}

To illustrate the shape of the coefficients with very different noise scenarios, the noise ASDs and phonon pulse template from two different datasets (A and B) from a single phonon channel of a SuperCDMS SNOLAB prototype detector are shown in figure~\ref{fig:G124_PSD_FIR} (a). The data were collected at an underground test facility. A noise environment similar to dataset A is what is expected when operations are going well, while dataset B describes the case of a more noisy environment, featuring significantly more noise spikes in the ASD above 100\,Hz. The OF FIR coefficients are derived separately for the two datasets and compared with a BF and a matched filter (MF) in
figure~\ref{fig:G124_PSD_FIR} (b). The matched filter coefficients resemble the shape of the signal template $A(t)$ \cite{matched_filter}.
The propagation of the noise spikes from dataset B to the OF FIR coefficients is clearly visible in the baseline oscillations surrounding the peak coefficients. Table~\ref{tab:filterPerformance} compares the resolution and response of the filter designs evaluated on the PSD and template of dataset A (and B). For comparability to the noise baseline, the FIR resolution is given in units of the standard deviation of the noise, $\sigma_{\rm n}$.

\begin{table}[b!]
    \centering
    \begin{tabular}{|c|c|c|c|c|}
        \hline
         Figure of Merit & Optimal Filter A & Optimal Filter B & Matched  Filter & Boxcar Filter\\
        \hline
        Resolution [$\sigma_{\rm n}$] & 0.39 (0.40) & 0.38 (0.39) & 0.52 (0.54) & 0.61 (0.64) \\
        Response & 2.37 & 2.31 & 2.93 & 3.49 \\
        $\frac{\rm{Response}}{\rm{Resolution}}$ [1/$\sigma_{\rm n}$] & 6.14 (5.90) & 6.10 (5.96) & 5.65 (5.39) & 5.68 (5.46) \\[0.2em]
        \hline
    \end{tabular}
    \caption{Resolution, response, and their ratio for different FIR filter designs evaluated on the power spectral density and pulse template of the SuperCDMS SNOLAB prototype dataset A. The evaluation was additionally repeated with dataset B with the results given in brackets. Optimal Filter A refers to the case when its coefficients were derived from dataset A, similarly for Optimal Filter B. For comparability to the noise baseline, the FIR resolution is given in units of the standard deviation of the noise, $\sigma_{\rm n}$. The differences between the optimal filter, matched filter, and boxcar filter are representative for a variety of different power spectral densities (PSDs).}
    \label{tab:filterPerformance}
\end{table}
The OF has the best resolution, whereas the BF has the best response. The ratio of response and resolution, an estimator for the signal-to-noise ratio, is best for the OF. Even when the OF is evaluated on a more noisy dataset it has not been derived from, it still performs better than the MF and BF. The filter performance is analyzed in more detail in section~\ref{sec:QualitativePerformance}.

\section{L1 Trigger Performance}
\label{sec:perf}
Given the predictions from section~\ref{sec:theo}, we evaluate the performance of the L1 trigger and the filter design in terms of their efficiency and noise rate at a given threshold. First, in section~\ref{sec:QualitativePerformance}, we demonstrate qualitatively the functioning and merit of the complex trigger architecture using a configuration that allows OF triggering while suppressing echo triggers by combining the OF with a BF in a separate trigger path. The quantitative performance of the different filter designs, OF, MF and BF, are then evaluated and compared in section~\ref{sec:QuantitativePerformance} using simulated detector noise conditions from a SuperCDMS SNOLAB prototype detector. We also compare how the OF performance changes when the noise conditions change.

\subsection{Qualitative Performance}
\label{sec:QualitativePerformance}

In this section the functionality of the L1 trigger architecture is demonstrated using test bench data and a trigger configuration that allows OF triggering while suppressing echo triggers by combining OF and BF information. 

\begin{figure}[!t]
     \centering
     \begin{subfigure}[b]{0.495\textwidth}
         \centering
         \includegraphics[width=\textwidth]{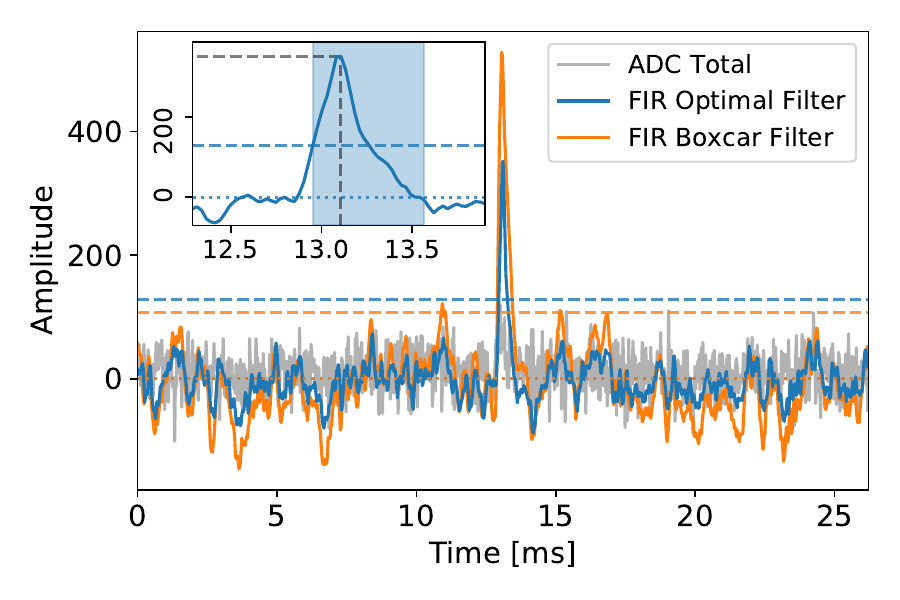}
         \caption{}
         \label{fig:FilteredTraceSmallamp}
     \end{subfigure}
     \hfill
     \begin{subfigure}[b]{0.495\textwidth}
         \centering
         \includegraphics[width=\textwidth]{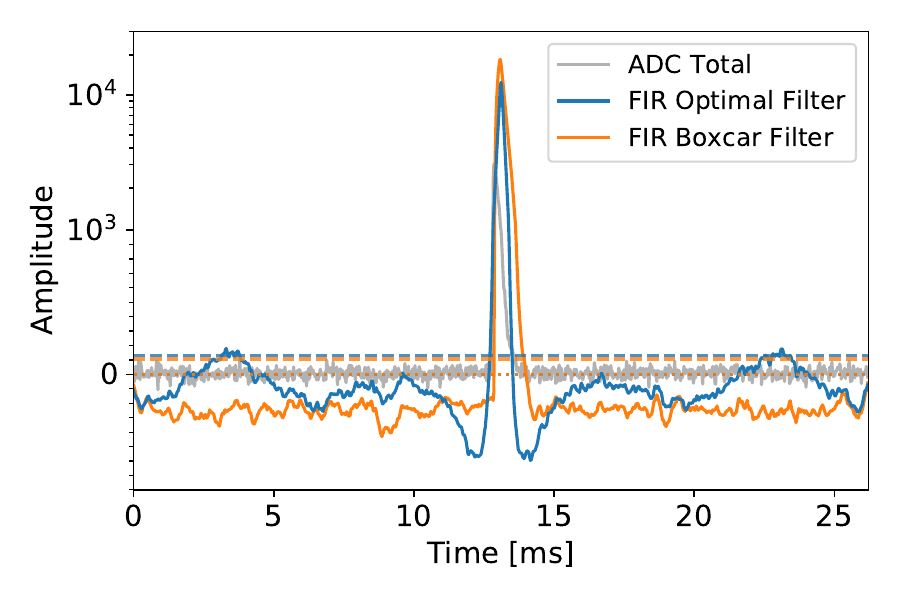}
         \caption{}
         \label{fig:FilteredTraceLargeamp}
     \end{subfigure}
        \caption{ADC readout traces and two simulated FIR outputs from an injected AWG pulse with an amplitude of 3 $\sigma_{\rm n}$ (a) and 100 $\sigma_{\rm n}$ (b) where the FIRs are configured as an optimal filter and a boxcar filter. The dashed and dotted lines mark the activation and deactivation thresholds. The blue shaded area in the inset of (a) marks the peak search window defined by the FIR filtered traces crossing above and falling below the activation and deactivation thresholds. The gray dashed lines indicate the final trigger peak amplitude and time stamp. The filtered trace of the optimal filter in (b) going above the activation threshold demonstrates the occurrence of echo triggers close to large pulses, the filtered trace of the boxcar filter does not show the same effect.}
        \label{fig:FilteredTraces}
\end{figure}

In the test bench setup, two DCRC phonon channels receive input pulses from an AWG. The AWG pulse shapes match the templates shown in figure~\ref{fig:TB_PSD_FIR} (a). The L1 trigger is configured with two trigger paths, which run in parallel an OF and a BF on the linear combination of the two input channels. The trigger activation (deactivation) thresholds seeded by the OF and BF trigger paths are set to 5 (0) $\sigma_{\rm{n}}^{\rm{OF}}$  and 2.5 (0) $\sigma_{\rm{n}}^{\rm{BF}}$  respectively. Figure~\ref{fig:FilteredTraces} shows the simulated FIR output response of the trigger paths to the physically received ADC signal from two different injected AWG pulse with an amplitude of 3 $\sigma_{\rm n}$ (a) and 100 $\sigma_{\rm n}$ (b).

By varying the amplitude of the input pulse we can separate the scenarios where there are no echo triggers and where there are. For the smaller injected signal the OF filtered trace exceeds the corresponding activation threshold (indicated by the dashed horizontal line) only in the peak region, whereas for the large injected signal the threshold is exceeded in two additional regions resulting in the previously mentioned echo triggers. By contrast, the BF activation threshold is set relatively low and is crossed several times by the BF filtered trace containing the small injected signal. In the presence of the large signal however, the negative BF FIR coefficients pull the BF filtered trace to negative FIR output values except for the peak region. This illustrates how the opposing behavior of OF and BF are utilized to take advantage of the superior OF performance while suppressing the echo triggers using the BF information. To implement this, a trigger logic module seeded by the OF trigger path is set to require an at-peak threshold logic bit of the OF threshold and a during-window threshold logic bit of the BF threshold. This way the final OF trigger path will only produce triggers if both the OF and BF threshold logic bits are high. The relatively low BF threshold ensures that the combined efficiency is not affected. 

\begin{figure}[!b]
    \centering
    \includegraphics[width=0.75\textwidth]{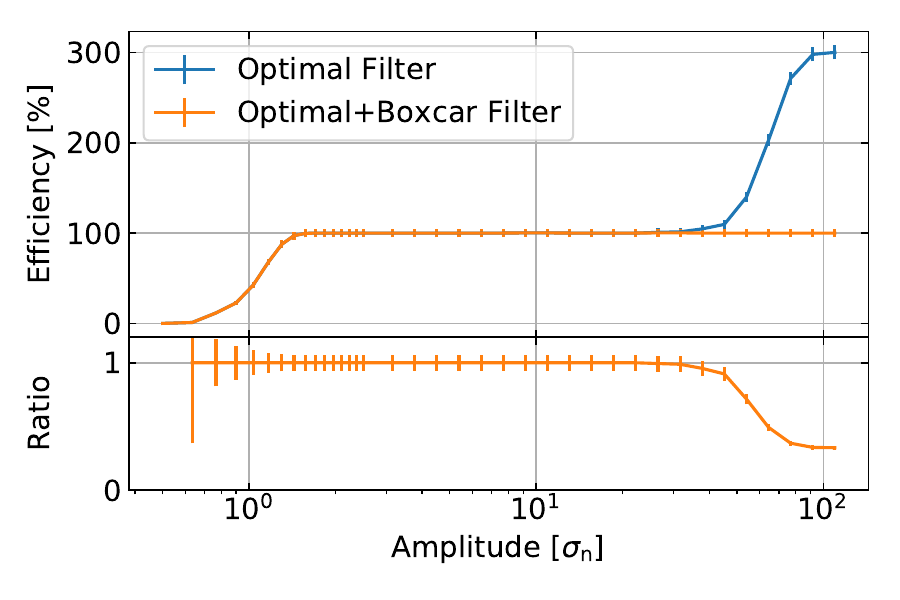}
    \caption{Efficiency turn-on curve as a function of injected AWG pulse amplitude for a simple OF trigger path and a combined trigger logic of an OF and a BF. An above-100\% efficiency corresponds to additional echo triggers found per pulse injected. The OF+BF combination reliably suppresses echo triggers without degrading the OF efficiency in the turn-on region.}
    \label{fig:efficiency_TB}
\end{figure}

To validate this functionality quantitatively multiple AWG pulses are injected at varying amplitudes and the trigger efficiency is measured as the number of received triggers divided by the number of injected signals. Figure \ref{fig:efficiency_TB} shows the efficiency as a function of injected pulse amplitude comparing a single OF configuration with the described OF+BF trigger logic. 
The efficiency turn-on starts at injected signal amplitudes of 0.6 $\sigma_{\rm n}$ and saturates at 1.6 $\sigma_{\rm n}$. For the single OF an additional echo trigger turn-on becomes dominant at 40 $\sigma_{\rm n}$ and saturates at 100 $\sigma_{\rm n}$. A 300\% efficiency corresponds to three triggers, two of which being echo triggers, found per pulse injected. In the OF+BF combination all echo triggers are suppressed and the efficiency turn-on is not degraded compared to a single OF configuration. Overall it is demonstrated that all trigger modules work as intended in a complex configuration on the hardware with external physical pulses. The trigger architecture can be configured elegantly to exploit information of multiple trigger paths, in this example, to suppress echo triggers without efficiency degradation.

\subsection{Quantitative Performance}
\label{sec:QuantitativePerformance}

The L1 trigger performance using the OF is determined in the presence of simulated detector noise and detector phonon pulses using the trigger simulation. The figures of merit are the efficiency turn-on curves, noise rates and the robustness towards changing noise. We note that since this is only for low energies, we will report our results for OF rather than the OF+BF for simplicity. For completeness, the results are compared to the MF and the BF.

\begin{figure}[!b]
     \centering
     \begin{subfigure}[b]{0.495\textwidth}
         \centering
         \includegraphics[width=\textwidth]{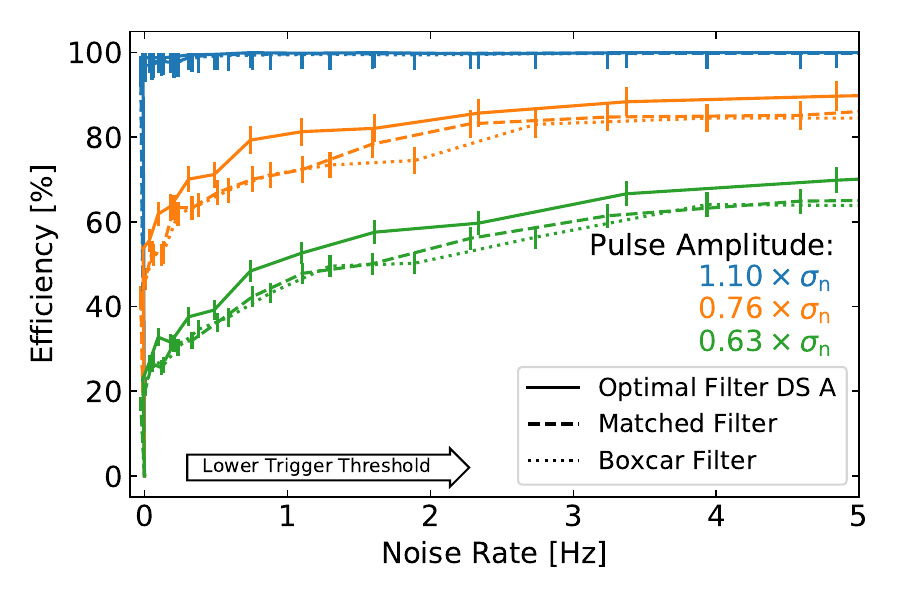}
         \caption{}
         \label{fig:SimEffVsNoiserate}
     \end{subfigure}
     \hfill
     \begin{subfigure}[b]{0.495\textwidth}
         \centering
         \includegraphics[width=\textwidth]{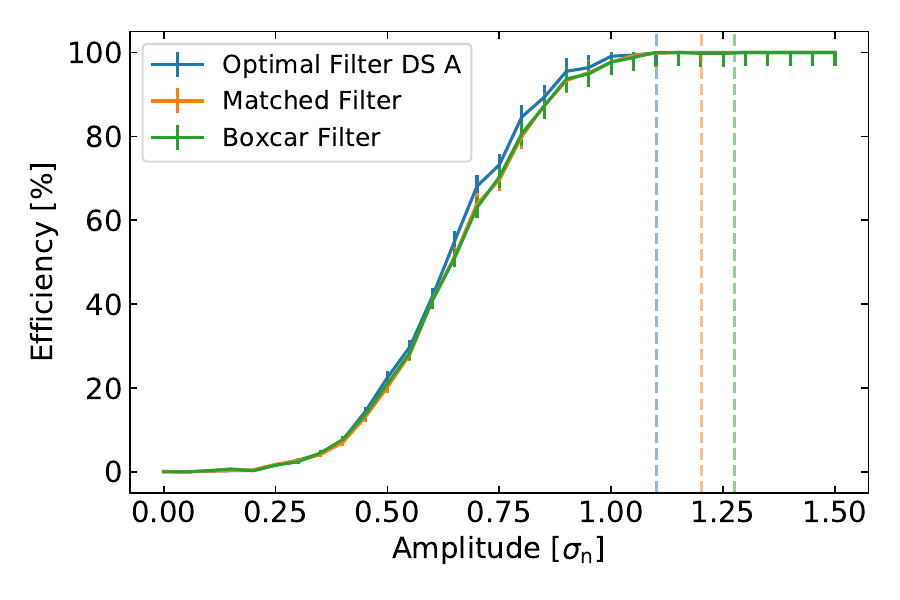}
         \caption{}
         \label{fig:SimEfficiency}
     \end{subfigure}
        \caption{(a) Trigger efficiency of different FIR filter designs as a function of noise rate for constant injected AWG pulse amplitudes. An increasing noise rate corresponds to a lower trigger threshold. (b) Trigger efficiency of different FIR filter designs as a function of injected AWG pulse amplitude at a trigger threshold corresponding to a 1\,Hz noise rate. The vertical dashed lines indicate the points of 99.9\% efficiency of the respective filters. The underlying noise conditions are simulated from a PSD taken from dataset DS A from the SuperCDMS SNOLAB prototype detector.}
        \label{fig:SimEfficiency_DSA}
\end{figure}

The PSDs and pulse templates from a single channel of dataset A from the SuperCDMS SNOLAB prototype detector shown in figure~\ref{fig:G124_PSD_FIR} (a) are used to create simulated pulses on a noisy baseline. An OF optimized to dataset A, an MF, and a BF with the respective FIR coefficients shown in figure~\ref{fig:G124_PSD_FIR} (b) are used to compare the trigger performance. The trigger activation threshold is varied and for each threshold data point the noise rate and the trigger efficiency are evaluated. The noise rate is taken as the average number of triggers per time in which the trigger was active without any signals being injected. The trigger efficiency is the number of triggered pulses per pulses injected at a fixed amplitude. The resulting measured trigger efficiency is plotted against the noise rate determined by the trigger threshold in figure~\ref{fig:SimEfficiency_DSA} (a). 
For a given noise rate, the OF is significantly more efficient than the MF and BF, independent of the injected pulse amplitude. The efficiency of MF and BF are similar. For the experiment operations, a threshold working point will be chosen with an acceptable noise rate. Assuming this working point at a noise rate of 1\,Hz, the trigger threshold can be fixed and a trigger turn-on curve can be determined by varying the injected pulse amplitude and measuring the trigger efficiency at each data point. The turn-on curves are shown in figure~\ref{fig:SimEfficiency_DSA} (b).

The efficiency turn-on starts at around 0.35 $\sigma_{\rm n}$ for all filter designs. In the upper region of the turn on, the OF performs better than the MF and BF, reaching its point of 99.9\% efficiency at 1.10 $\sigma_{\rm n}$ compared to 1.20 $\sigma_{\rm n}$ and 1.27 $\sigma_{\rm n}$ for the MF and BF respectively.

\begin{figure}
     \centering
     \begin{subfigure}[b]{0.495\textwidth}
         \centering
         \includegraphics[width=\textwidth]{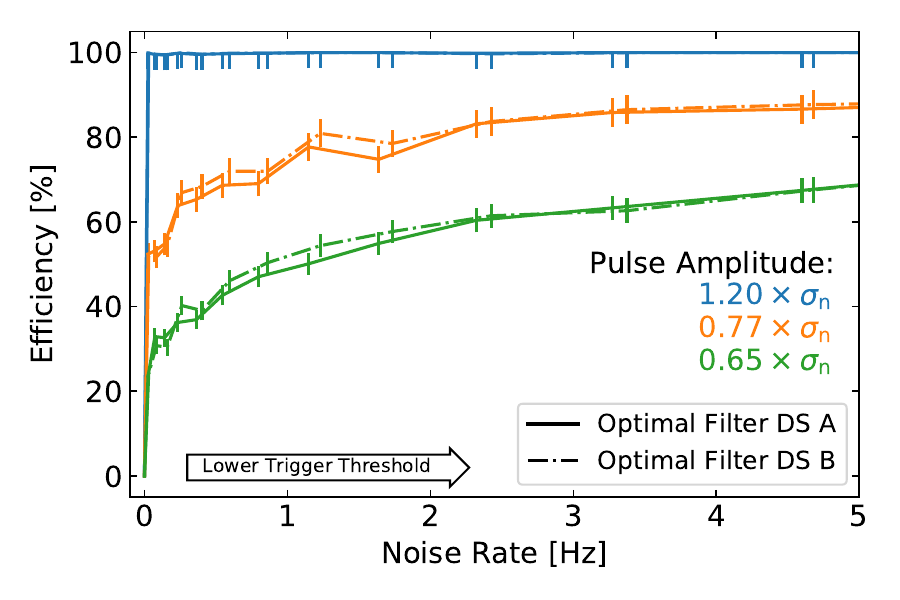}
         \caption{}
         \label{fig:SimEffVsNoiserate_DSB}
     \end{subfigure}
     \hfill
     \begin{subfigure}[b]{0.495\textwidth}
         \centering
         \includegraphics[width=\textwidth]{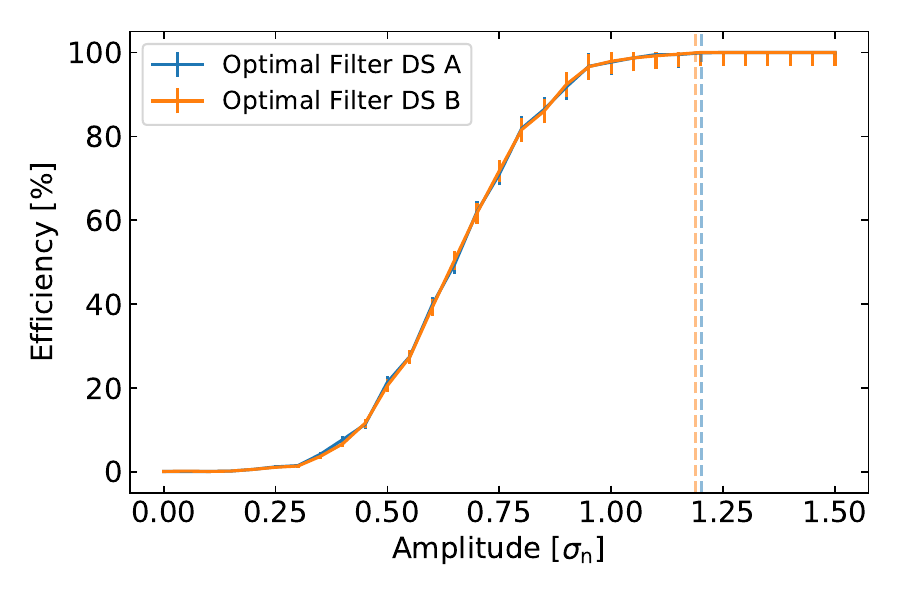}
         \caption{}
         \label{subfig:SimEfficiency_DSB}
     \end{subfigure}
        \caption{(a) Comparison of the trigger efficiency of two optimal filters, respectively optimized to dataset DS A and DS B, as a function of noise rate determined by the trigger threshold for constant injected AWG pulse amplitudes. (b) Trigger efficiency of the two optimal filters as a function of injected AWG pulse amplitude at a trigger threshold corresponding to a 1\,Hz noise rate. The underlying noise conditions are simulated from a PSD taken from dataset B from the SuperCDMS SNOLAB prototype detector. The points of 99.9\,\% efficiency are at 1.20 $\sigma_{\rm n}$ and 1.19 $\sigma_{\rm n}$ for the OF optimized to dataset DS B and DS A.}
        \label{fig:SimEfficiency_DSB}
\end{figure}

During a science exposure period it is possible for the noise conditions to change, which the OF trigger ideally should be robust to. We investigate how the filter performance changes when the noise conditions change with and without a re-optimization of the OF coefficients. For this, the performance of the two OFs, optimized to datasets A and B from the SuperCDMS SNOLAB prototype detector, with respective FIR coefficients shown in figure~\ref{fig:G124_PSD_FIR} (b), are compared based on simulated noise traces and pulses from dataset B. The PSD from dataset B features multiple noise spikes which could realistically occur occasionally during a data taking period. Figure~\ref{fig:SimEfficiency_DSB} illustrates the efficiency as a function of noise rate and injected pulse amplitude and demonstrates that the efficiency degrades only slightly if the OF is not re-optimized to the changed noise conditions\footnote{The vice versa case where the OFs are applied to dataset A leads to a similar difference in efficiency.}. Thus the OF shows robustness against noise fluctuations indicating that re-optimization is desirable but not expected to be crucial.

\section{Conclusion}
\label{sec:conclusion}
The SuperCDMS L1 trigger system is the hardware-based first stage of the three-staged SuperCDMS trigger system. Embedded in a multi-modular architecture on an FPGA on custom DCRCs, it is designed to provide efficient, dead-time-free triggering at a threshold that is as low as possible. In four independently configurable trigger paths, digitized input traces from up to 16 channels are downsampled, linearly combined, and FIR-filtered. For each trigger path the threshold, peak amplitude, and peak timestamp information are extracted from the filtered traces. Based on this combined information across the trigger paths, a flexible trigger logic takes the trigger decisions and stores the final trigger primitive information into a FIFO buffer. The expected functioning of individual trigger modules and the system as a whole is validated using logic simulation software and a software implementation of the trigger. The default configuration of the FIR filter is a time-domain OF which, in typical noise conditions, has a better response to resolution ratio than an MF and a BF, but can cause spurious echo triggers in the presence of large pulses. The functioning and merit of the complex trigger architecture is demonstrated qualitatively in a configuration that suppresses echo triggers without efficiency degradation via the combination of an OF and a BF in two trigger paths. Using the trigger simulation software, the quantitative performance is determined via the measurement of the trigger efficiency and noise rate for different filter designs, trigger thresholds, pulse amplitudes, and expected noise conditions. The OF proves to be the superior configuration with an efficiency of 99.9\% for signal amplitudes of 1.1 $\sigma_{\rm n}$ in typical noise conditions and a high degree of robustness towards potentially occurring noise spikes. For the beginning of data taking with SuperCDMS SNOLAB expected in 2023/2024, the L1 trigger is ready and well-prepared to efficiently trigger potential dark matter events at the high and low energy frontier ensuring a rich and diverse physics program.

\acknowledgments

We would like to thank Caleb Fink for helping with the initial commissioning and testing of the level-1 trigger and Lucas Scholl for characterizing echo triggers in test bench data. We are grateful to Stefan Zatschler for comments on a draft version of this manuscript. We gratefully acknowledge support from the U.S. Department of Energy (DOE) Office of High Energy Physics under the award numbers DE-SC0018981 and DE-SC0015657, from the National Science Foundation (NSF), from NSERC Canada and the Arthur~B.~McDonald Institute, and from the Deutsche Forschungsgemeinschaft (DFG, German Research Foundation) under the Emmy Noether Grant No. 420484612, and under Germany's Excellence Strategy - EXC 2121 ``Quantum Universe'' -  390833306.  Fermilab is operated by Fermi Research Alliance, LLC, for the DOE under Contract No. DE-AC02-37407CH11359. Pacific Northwest National Laboratory (PNNL) is operated by Battelle Memorial Institute for the DOE under Contract No. DE-AC05-76RL01830. SLAC is operated for the DOE under Contract No. DEAC02-76SF00515. Lawrence Berkeley National Laboratory (LBNL) is operated for the DOE under Contract No. DE-AC02-05CH11231. We thank the Mitchell Institute for Fundamental Physics and Astronomy at Texas A\&M University for providing support for this work.

\bibliographystyle{iopart-num}
\bibliography{references}{}

\providecommand{\newblock}{}
\begin{thebibliography}{10}
\expandafter\ifx\csname url\endcsname\relax
  \def\url#1{{\tt #1}}\fi
\expandafter\ifx\csname urlprefix\endcsname\relax\def\urlprefix{URL }\fi
\providecommand{\eprint}[2][]{\url{#2}}

\bibitem{Agnese_2017}
Agnese R {\em et~al.\/} (SuperCDMS) 2017 {\em Physical Review D\/} {\bf 95}
  ISSN 2470-0029 \urlprefix\url{http://dx.doi.org/10.1103/PhysRevD.95.082002}

\bibitem{Neganov:1985khw}
Neganov B~S and Trofimov V~N 1985 {\em Otkryt. Izobret.\/} {\bf 146} 215

\bibitem{Luke1988VoltageassistedCI}
Luke P~N 1988 {\em Journal of Applied Physics\/} {\bf 64} 6858--6860

\bibitem{cycloneIV}
{Intel Altera FPGA} 2016 {Cyclone® IV Device Handbook} Tech. Rep. CYIV-5V1-2.2

\bibitem{Avalon-spec}
{Intel FPGA} 2021 Avalon® interface specifications Tech. Rep. MNL-AVABUSREF

\bibitem{CIC_filter}
Hogenauer E 1981 {\em IEEE Transactions on Acoustics, Speech, and Signal
  Processing\/} {\bf 29} 155--162

\bibitem{z-transform}
Jury E~I 1973 {\em Theory and application of the z-transform method\/}
  (Huntington, N.Y., R.E. Krieger Pub. Co.) ISBN 0-88275-122-0

\bibitem{fir_stability}
Oshana R 2012 Chapter 7 - overview of dsp algorithms {\em DSP for Embedded and
  Real-Time Systems\/} ed Oshana R (Oxford: Newnes) pp 113--131 ISBN
  978-0-12-386535-9
  \urlprefix\url{https://www.sciencedirect.com/science/article/pii/B978012386535900007X}

\bibitem{Quartus-ModelSim}
{Intel Altera FPGA} 2015 {Quartus® II Handbook (ModelSim*-Intel® FPGA Edition
  Software)} Tech. Rep. QII5V1

\bibitem{matched}
Smith S~W 1997 {\em The Scientist and Engineer's Guide to Digital Signal
  Processing\/} (USA: California Technical Publishing) ISBN 0966017633

\bibitem{boxcar}
Roscoe A~J and Blair S~M 2016 Choice and properties of adaptive and tunable
  digital boxcar (moving average) filters for power systems and other signal
  processing applications {\em 2016 IEEE International Workshop on Applied
  Measurements for Power Systems (AMPS)\/} pp 1--6

\bibitem{OF_gatti}
Gatti E and Manfredi P~F 1986 {\em La Rivista del Nuovo Cimento (1978-1999)\/}
  {\bf 9} 1--146

\bibitem{matched_filter}
Turin G 1960 {\em IRE Transactions on Information Theory\/} {\bf 6} 311--329

\end{thebibliography}

\end{document}